\documentclass{icrc29}
\usepackage{graphicx,amssymb,amsmath,times}
\begin{document}
%Title of paper
\title[Search for Pulsed VHE Gamma-ray Emission]{Search for Pulsed VHE $\gamma$-ray Emission from
Young Pulsars with H.E.S.S.}
\author[A. Konopelko et al.] {A. Konopelko$^{a,b}$, P. Chadwick$^c$, T. Eifert$^a$,
        T. Lohse$^a$, A. Noutsos$^c$,
        S. Rayner$^c$,   \newauthor F. Schmidt$^a$, U. Schwanke$^a$,
        Ch. Stegmann$^a$, for the H.E.S.S. collaboration \\
        (a) Institute of Physics, Humboldt-University for Berlin, D-12489 Berlin, Germany\\
        (b) Max-Planck-Institute of Nuclear Physics, Postfach
        103980, Heidelberg, Germany\\
        (c) Department of Physics, University of Durham, Rochester Building,
            Science Laboratories,\\ South Road, Durham DH1 3LE
        }
\presenter{Presenter: A. Konopelko
(Alexander.Konopelko@mpi-hd.mpg.de), \
ger-konopelko-A-abs4-og22-poster}

\maketitle

\begin{abstract}
Pulsars are commonly regarded as highly magnetized neutron stars,
rotating up to several hundred times per second. Over 1,500 radio
pulsars have been found so far, about 70 of which are X-ray
pulsars, but only a handful have been observed in $\gamma$ rays.
This high-energy emission is believed to be produced by the
electrons accelerated to TeV energies in the pulsar magnetosphere,
resulting in cascades of secondary particles. The H.E.S.S.
experiment is a system of four imaging atmospheric Cherenkov
telescopes in Namibia, which is sensitive to $\gamma$ rays above
100 GeV. Three young pulsars, Crab, Vela, and PSR B1706-44 have
been observed with H.E.S.S. The results of the search for pulsed
emission for these targets, and the constraints on the theories of
pulsed very high energy $\gamma$-ray emission, are summarized
here.
\end{abstract}

\vspace*{-2mm}
\section{Introduction}
\vspace*{-1mm}

The EGRET telescope on the Compton Gamma Ray Observatory (CGRO)
identified at least six young spin-powered $\gamma$-ray pulsars
\cite{Thompson00} at energies up to 20~GeV. A search for pulsed
emission above 20~GeV remains to be successful for ground-based
Cherenkov instruments. Despite a reduction of the energy threshold
down to 60~GeV \cite{deNaurois02}, achieved with the the
non-imaging telescopes, no evidence of a pulsed signal has been
seen so far. It is widely accepted that this unexplored energy
region is vital to address the long-standing question of how and
where high energy emission emerges from the pulsar.

{\it Polar cap} models (for review of the models see
\cite{Harding00}) assume that particles are accelerated right
above the neutron star surface and that $\gamma$ rays emanate from
curvature radiation or inverse Compton induced pair cascades in a
strong magnetic field. On the other hand {\it outer gap} models
consider particle acceleration which occurs along null charge
surfaces in the outer magnetosphere where $\gamma$-rays result
from photon-photon pair production-induced cascades. These two
types of basic models make rather different predictions,
particularly of the spectral characteristics of pulsed
$\gamma$-ray emission. Detection and study of this high energy
emission, which is closely tied to the primary population of
radiating particles, seems to be a good discriminant between
pulsar models.

The {\it High Energy Stereoscopic System} (H.E.S.S.) of four
imaging atmospheric Cherenkov telescopes with a low energy
threshold of 100~GeV at zenith and below 1\% Crab flux sensitivity
for long exposures \cite{Hinton04} was used for observations of
three young pulsars: the Crab and Vela pulsars, and PSR~B1706-44,
characterized by their high-ranking position on a list of pulsars
ordered by the parameter $\dot{E}/d^2$, which ultimately
determines pulsar luminosity at high energies for a given pulsar
age. Here $\dot{E}$ is the spin-down energy loss rate, and $d$ is
the distance determined from the dispersion measure to the object.
The spectra of the EGRET-detected pulsars can be described quite
well by simple power laws with spectral indices in the range
1.39-2.07 \cite{Nolan96}. Some of the spectra show clear evidence
for a drop in flux at high energy, above few GeV, whereas others
have rather large uncertainties in the 4-10~GeV band, which
prevent the clear identification of a similar cutoff. The measured
Crab spectrum is a straight power law. Extrapolation of EGRET
spectra into the dynamic range of H.E.S.S. suggested that high
energy pulsed emission might be observable from these objects
within a reasonable exposure time \cite{deJager01}. The outer gap
model also supplements pulsed TeV emission via inverse Compton
scattering by gap-accelerated particles. The most recent outer gap
models \cite{Hirotani01} have TeV $\gamma$-ray fluxes for the Vela
pulsar which should be detectable with H.E.S.S.

H.E.S.S. is an array of four atmospheric imaging Cherenkov
telescopes, each with $\rm 107\,m^2$ of mirror area and equipped
with a 960 photo-multiplier tube camera \cite{Hinton04}.
Telescopes are operated in a stereoscopic mode with a system
trigger, requiring at least two telescopes to provide images of
each individual atmospheric shower in Cherenkov light. H.E.S.S.
has large field of view of $5^\circ$ diameter. The angular
resolution for individual $\gamma$ rays is better than
$0.1^\circ$. It allows a very good source localization accuracy of
30$''$ for relatively faint sources, which is important for
point-like source (e.g. pulsars) identification.

\vspace*{-3mm}
\section{Data Sample \& Analysis}
\vspace*{-1mm}

A substantial fraction of the data on young pulsars was taken
during the construction phase of the H.E.S.S. array, when only two
or three telescopes were available. Short summary of data is given
in Table~\ref{data}. The Crab pulsar, located in the northern sky,
can be observed with H.E.S.S. only at a rather large average
zenith angle ($<\Theta>$) and consequently above rather a high
energy threshold ($\rm E_{th}$), whereas Vela and PSR B1706-44 can
be seen with H.E.S.S. at much higher elevations. Most of the data
were taken in so-called {\it wobble} source-tracking mode, which
is optimal for observations of a point-like source. A few
additional hours of observations of PSR~B1706-44 have been
extracted from the long scan of the galactic plane \cite{galscan}.

\begin{table}[t]   %[H] add [H] placement to break table across pages
\caption{\label{data} Summary of data samples} \vspace*{1mm}
 \centering
%\begin{center}
\begin{tabular}{||c|c|c|c|c|c||} \hline \hline
~Source~ & ~Setup~ & ~Obs. period~ & ~t [hr]~ & ~$\rm E_{th}$
[GeV]~ & ~$<\Theta>$ [deg]~ \\ \hline \hline Crab & 3 Tel. &
Oct'03 & 4.0 & 350 & 47 \\ \hline Vela & 4 Tel. & Jan-March'04 &
12.6 & 235 & 33 \\ \hline PSR B1706-44 & 2 Tel. \& & May-July'03
\& &
14.4 + & 255 & 27 \\
 & 4 Tel. & June-July'04 & 2.2 & & \\ \hline
\end{tabular}
%\end{center}
\vspace*{-2mm}
\end{table}

Prior to applying analysis cuts, data were selected for adequate
recorded image quality, i.e. by applying the generic requirements
of a minimum image amplitude (50 photoelectrons) and a maximum
distance of the image's centroid from the camera center (35 mrad).
Each accepted event was also required to comprise at least two
images of adequate quality. The imaging analysis of the H.E.S.S.
data is based on the reconstruction of the shower direction for
each individual event, and joint parametrization of the shape of
the Cherenkov light flash from an individual shower using
multiple-telescope approach. Data were analyzed by the standard
{\it directional} cut on $\theta^2$, where $\theta$ is the angular
distance between the actual source position on sky and the
reconstructed one. In addition data were analyzed by the image
{\it shape} parameters of the mean scaled Width and Length
\cite{pks2155} (see Table~\ref{cuts}). The analysis cuts were
optimized on the Monte Carlo simulated events for the zenith angle
range covered in observations of a particular source, and the
system configuration used (see Table~\ref{data}). Details of the
methods used to estimate the effective areas, the energy
threshold, and the energy for each recorded event are given
elsewhere \cite{pks2155}. The upper limits on pulsed $\gamma$-ray
emission were calculated here using a {\it model-independent}
approach \cite{psr1706}, which exploits the energy information
provided for each individual event.

The arrival times of the recorded events were registered by a
Global Positioning System (GPS) clock with an absolute accuracy of
about 1~$\mu \rm s$. The arrival times were recalculated to the
solar system barycenter (SSB) by utilizing the JPL DE200 ephemeris
\cite{Standish82}. The universal time as measured at the H.E.S.S.
site was converted to the SSB arrival time (TDB). The corrected
arrival times were transformed into phases of the radio pulse
period using publicly available contemporary pulsar ephemerides
obtained from radio observatories, which monitor the pulsars
observed with H.E.S.S. on a monthly basis. The resulting phasogram
for each of observed pulsars was a subject to a number of
statistical tests, i.e. $\chi^2$, $Z^2_m$, $H$ (for details see
\cite{okkie94}), which allow the assessment of the significance of
a pulsed signal for a variety of possible pulse profiles. A low
chance probability of a uniform phasogram, i.e. less than
$10^{-3}$, derived from any of these tests might indicate a pulsed
signal.

\begin{table}[t]
\caption{\label{cuts} The $\gamma$-ray selection criteria}
\vspace*{1mm} \centering
%\begin{center}
\begin{tabular}{||c|c|c|c||} \hline \hline
 & ~Crab~ & ~Vela~ & ~PSR B1706-44~ \\ \hline \hline
$\theta^2,\,\,\rm deg^2$  & $<$0.055 & $<$0.02 & $<$0.02 \\
mean scaled Width  & $<$1.0 & $<$0.9 & $<$1.1 \\
means scaled Length  & $<$1.4 & $<$1.3 & $<$1.3 \\ \hline
\end{tabular}
%\end{center}
\vspace*{-4mm}
\end{table}

In order to verify the periodic analysis procedure, observations
of the optical emission from Crab pulsar have been performed with
a single stand-alone H.E.S.S. telescope. The optical data were
folded to produce the phases using full chain of the H.E.S.S.
periodic analysis tools. The distinct double-peaked optical light
curve of the Crab pulsar at the correct phase was clearly
resolved, which validates the periodic analysis \cite{optical}.

\vspace*{-2mm}
\section{Results \& Discussion}
\vspace*{-1mm}

\begin{figure}[t]
\centering
%\begin{center}
\includegraphics*[width=0.58\textwidth,angle=0,clip]{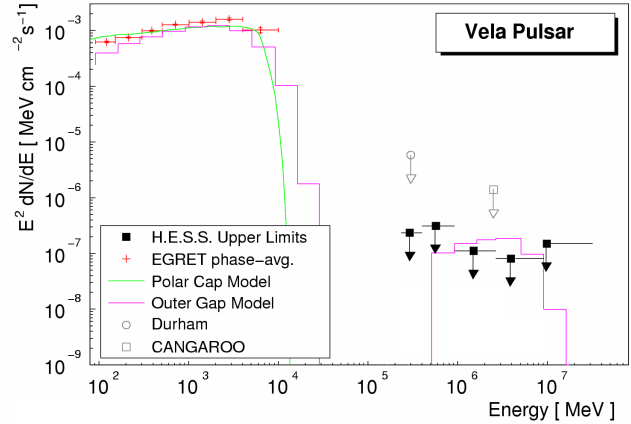}
\vspace*{-2mm} \caption{\label {fig1} The energy spectrum of the
pulsed emission from the Vela pulsar as measured by EGRET together
with the H.E.S.S. upper limits. Predictions from the polar cap
(solid curve) \cite{Harding00} and outer gap model (histogram)
\cite{Hirotani01} are also shown. Other TeV upper limits are given
for Durham \cite{durham} and CANGAROO \cite{cangaroo}
experiments.}
%\end{center}
\vspace*{-2mm}
\end{figure}

Data taken with the H.E.S.S. system of four imaging atmospheric
Cherenkov telescopes have been used to search for pulsed
$\gamma$-ray emission from the Crab, Vela pulsars, and
PSR~B1706-44. No pulsed emission was found at the radio period for
any of these pulsars, and corresponding upper limits on integral
as well as on differential flux have been derived (see
Table~\ref{results}, Figure~\ref{fig1}). The upper limits have
been calculated for the phase regions selected according to the
EGRET peak areas.

To extract the low energy events exclusively, we analyzed the data
using a set of tightly adjusted cuts. In particular the total
image amplitude must be less that 100 photoelectrons and maximum
distance must not exceed 18~mrad. Such analysis resulted in
differential upper limits (99\%) of $\rm dN/dE(232\pm 51\,\,GeV) =
3.94\times 10^{-11}\,cm^{-2}s^{-1}TeV^{-1}$ for the Crab pulsar,
and $\rm dN/dE(75\pm 12\,\,GeV) = 5.2\times
10^{-10}\,cm^{-2}s^{-1}TeV^{-1}$ for PSR~B1706-44. Note that a
disadvantage of this analysis is that it drastically suppresses
the $\gamma$-ray acceptance at higher energies.

We modelled the pulsed $\gamma$-ray spectrum using the following
form $dN/dE = C E^{-\gamma}e^{-E/E_c}, \label{fit}$ where $C$ is
the EGRET measured normalization constant, $\gamma$ is the index
of the known EGRET power-law spectrum, and $E_c$ is the maximal
cutoff energy, which is consistent with the integral upper limit
given in Table~\ref{results}. The cutoff energy can be constrained
as $\int_{E_{th}}^\infty (dN/dE)dE \leq F(>E_{th})$.
\begin{table}[t]
\caption{\label{results} The H.E.S.S. upper limits (99\%) and
parameters of the EGRET spectral fit for three young pulsars}
\vspace*{1mm} \centering
%\begin{center}
\begin{tabular}{||c|c|c|c|l||} \hline \hline
 & ~$\gamma$~ & ~$\rm E_c$ [GeV]~ & ~Phase region~  &
 ~Upper limit [$\rm cm^{-2}s^{-1}$]~ \\ \hline \hline
Crab & 2.05 & 117 & [0.32,0.42]\&[0.94,0.04] & $\rm
F(>350\,GeV)\,<\, 4.67\times 10^{-12}$  \\
Vela & 1.62 & 26.5 & [0,0.13]\&[0.5,0.57] & $\rm
F(>235\,GeV)\,<\,7.17\times 10^{-13}$ \\
PSR~B1706-44 & 2.25 & 71 & [0.24,0.5]& $\rm
F(>255\,GeV)\,<\,1.06\times 10^{-12}$ \\ \hline
\end{tabular}
%\end{center}
\vspace*{-2mm}
\end{table}
Derived cutoff energies for three young pulsars observed with
H.E.S.S. are given in Table~\ref{results}. The stringent integral
upper limits above 200-300~GeV reported here for three young
pulsars after rather limited exposures with H.E.S.S. appear to be
still above the predictions by the polar cap and outer gap models
at these energies. Therefore, these upper limits cannot be used to
discriminate between two competing models. However, they can
severely restrain the luminosity of pulsed $\gamma$-ray emission.
In addition, for the Vela pulsar the outer gap model predicts a
rather flux emission via inverse Compton scattering at TeV
energies \cite{Hirotani01}, which is inconsistent with the
model-independent H.E.S.S. upper limits reported here. The inverse
Compton flux level depends on the emission spectrum mainly in the
infra-red band, which is difficult to measure in most pulsars.
Thus the H.E.S.S. upper limits in particular for Vela pulsar
constrain the density of local soft photon field in the gap.
% and ultimately force the revision of the outer gap models in
%order to comply with the H.E.S.S. upper limits at TeV energies
%presented here.

{\bf Acknowledgement.} {\small The support of the Namibian
authorities and of the University of Namibia in facilitating the
construction and operation of H.E.S.S. is gratefully acknowledged,
as is the support by the German Ministry for Education and
Research (BMBF), the Max Planck Society, the French Ministry for
Research, the CNRS-IN2P3 and the Astroparticle Interdisciplinary
Programme of the CNRS, the U.K. Particle Physics and Astronomy
Research Council (PPARC), the IPNP of the Charles University, the
South African Department of Science and Technology and National
Research Foundation, and by the University of Namibia. We
appreciate the excellent work of the technical support staff in
Berlin, Durham, Hamburg, Heidelberg, Palaiseau, Paris, Saclay, and
in Namibia in the construction and operation of the equipment.}

\end{document}